\documentclass[aps,prb,twocolumn,superscriptaddress]{revtex4-2}
\usepackage{amssymb,bm,graphicx,ulem,hyperref,amsmath,mathdots,xcolor}

\begin{document}

	\author{I. Chestnov}
\affiliation{Westlake University, School of Science, 18 Shilongshan Road, Hangzhou 310024, Zhejiang Province, China}
\affiliation{Westlake Institute for Advanced Study, Institute of Natural Sciences, 18 Shilongshan Road, Hangzhou 310024, Zhejiang Province, China}
\affiliation{Vladimir State University, Gorkii St. 87, 600000, Vladimir, Russia}
	\author{A. Yulin}
	\affiliation{Faculty of Physics, ITMO University, 197101, St. Petersburg, Russia}
	\author{I. A. Shelykh}
	\affiliation{Science Institute, University of Iceland, Dunhagi 3, IS-107, Reykjavik, Iceland}
	\affiliation{Faculty of Physics, ITMO University, 197101, St. Petersburg, Russia}
	\author{A. Kavokin}
\affiliation{Westlake University, School of Science, 18 Shilongshan Road, Hangzhou 310024, Zhejiang Province, China}
\affiliation{Westlake Institute for Advanced Study, Institute of Natural Sciences, 18 Shilongshan Road, Hangzhou 310024, Zhejiang Province, China}
\affiliation{Russian Quantum Center, Skolkovo, Moscow 143025, Russia}
	
	\title{Dissipative Josephson vortices in annular polariton fluids}

\begin{abstract}
We consider two concentric rings formed by bosonic condensates of exciton-polaritons. A circular superfluid flow of polaritons in one of the rings can be manipulated by acting upon the second annular polariton condensate. The complex coupling between the rings with different topological charges triggers nucleation of stable Josephson vortices (JVs) which are revealed as topological defects of the angular dependence of the relative phase between rings.   Being dependent on the coupling strength, the structure of the JV governs the difference of the mean angular momenta of the inner and the outer rings.  At the vanishing coupling the condensates rotate independently demonstrating no correlations of their winding numbers. At the moderate coupling, the interaction between two condensates tends to equalize their mean angular  momenta  despite of the mismatch of the winding numbers demonstrating the phenomenology of a drag effect. Above the critical coupling strength the synchronous rotation is established via the phase slip events.	
\end{abstract}
	
	\maketitle
	
\section{Introduction}

If two coherent quantum systems are separated by a weak link, their mutual coherence is sustained due to the tunnelling across the link. This phenomenon known as Josephson effect was observed on various physical platforms including superconductors \cite{barone1982}, liquid helium \cite{sato2011} and Bose-Einstein condensates \cite{levy2007}. A supercurrent flowing through the junction  represents a striking  manifestation of a quantum interference at macroscopic scale. Being an exceptional phenomenon at equilibrium, the internal currents
are intrinsic to open systems which are pumped from outside. In this case, the current originates from the local gain-dissipation imbalance and serves for its compensation. Its existence reflects an intrinsic tendency of the system to maintain the detailed dynamical balance characteristic to the stationary regime.

An important example of a coherent driven-dissipative system is a bosonic condensate of exciton polaritons. Polaritons are hybrid quasiparticles which appear in the strong light-matter coupling regime in semiconductor crystals and micro-structures. In low-dimensional semiconductor microcavities excited by optical pumping or electrically injected,  polaritons undergo a transition to the condensed state which demonstrates collective coherent phenomena typical to a superfluid \cite{KavokinMicrocavities,Carusotto2013}. The internal polariton currents stem naturally from the spatial inhomogeneity of the external pump and has a crucial impact on the polariton fluid. Being pointed from the region with the excess of gain to where it is scarce, the persistent flow of polaritons supports peculiar density patterns \cite{Wouters2008} such as dissipative solitons \cite{Ostrovskaya2012} and assists a self-trapping effect \cite{Chestnov2018}.  Besides, the global polariton current can appear spontaneously even at the homogeneous pumping due to the intrinsic instability of the current-less state \cite{Nalitov2017,Chestnov2016}. In an annular geometry, the local currents stemming from the chiral-symmetry breaking gain distribution 
allow for winding up polariton superfluid \cite{Dall2014,Sedov2021}.

If two polariton condensates are placed in close proximity, they form a Josephson junction \cite{Lagoudakis2010,Abbarchi2013}. As a result in the driven-dissipative systems the coupling typically acquires a dissipative component \cite{Aleiner2012}.  In contrast to the conventinonal Josephson coupling which corresponds to the real tunnelling amplitude, the dissipative coupling affects the decay rates of the interacting systems rather than their energies. The complex coupling results in the phase locking effect which was investigated with the individual polariton Josephson junctions \cite{Ohadi2016} as well as in a multi-condensate array, in the polariton graph \cite{Berloff2017,Lagoudakis2017} and lattice \cite{Ohadi2018,Topfer2021} configurations.  
Manipulation of the pump geometry and position offers a versatile tool for the precise tuning of the coupling strength \cite{Alyatkin2020,Cherotchenko2021}  which 
allows exploiting polariton junctions as building blocks of the quantum optimizers and for simulation of the complex spin problems \cite{kalinin2020}.

Considering the Josephson phenomenon one should distinguish between short (point-like) and long junctions. A short junction confines the supercurrent in a transverse direction suppressing the flow along the junction interface. Short junctions are typically used in superconducting qubit circuits \cite{Krantz2019} and liquid helium gyroscopes \cite{sato2011}. Two spatially separated spot-like polariton condensates  also form the short junction \cite{Ohadi2016}.

In long junctions the interface extends beyond the Josephson penetration depth \cite{barone1982}, so that the transversal dynamics of the Josephson current becomes essential. In this case, the current can twist into a closed loop forming the Josephson vortex (JV). These states also known as fluxons play a crucial role in the physics of supercodunctors \cite{wallraff2003} due to their ability of trapping magnetic flux quantum. Besides JVs can emerge between two elongated atomic condensates \cite{Kaurov2005,Kaurov2006,Brand2009}. Recently, JVs were observed in a polariton superfluid. In Ref.~\cite{Caputo2019} they appear as the topological excitations in the phase twisting region imposed by two phase-locking laser beams.

In spite of the progress of the experimental studies the understanding of the physics of the driven-dissipative JVs is still far from being achieved. In this paper, we study the formation of the JVs between two polariton condensates separated by long interface. We focus on the annular geometry and consider specifically two concentric ring-shaped polariton condensates. A similar configuration was studied in superconducting \cite{davidson1985,price2010} and atomic \cite{Brand2009} systems, where spontaneous nucleation of JVs in the annular junction was employed for testing of the Kibble-Zurek mechanism \cite{Su2013}. In contrast to the superconducting and atomic system, the polariton system under study is a driven-dissipative condensate out of thermal equilibrium. 

Out of equilibrium, the coexistence of the intrinsic internal and the Josephson currents as well as the entirely complex coupling parameter are expected to have a crucial impact on the polariton circulation in the double-ring geometry. Formation of the annular Josephson junction between two concentric polariton rings was investigated in Ref.~\cite{MunozMateo2020}. The stability of the uniform solutions with no circular currents in both rings was described, while the nucleation of JVs remained out of consideration. The excitation of identical circular currents in  both inner and outer polariton rings was predicted in \cite{Barkhausen2021}.

In this paper, we demonstrate that the rotation of polariton rings with different winding numbers is also possible. It is assisted by the trapping of the JV in the annular junction. The JVs are excited spontaneously during the condensate formation. Due to the interplay between internal currents and coherent tunneling, the relative rotation velocity of the rings can be tuned \textit{continuously}. In the presence of JV, both the inner and the outer condensates carry fractional angular momenta  whose difference is governed by the spatial structure of the JV. Thus the interaction between the ring-shaped condensates demonstrates phenomenology of the drag effect. Besides we predict the existence of the critical coupling strength above which the coherent coupling between the condensates forces them to rotate synchronously. This regime is established via the phase slip events which correspond  to the annihilation of the JVs.

\section{The model}
\subsection{The annular Josephson junction between the concentric ring-shaped polariton condensates}

We consider exciton polaritons condensates excited nonresonantly. The optical pump creates incoherent exctions which relax their energy and momenta feeding the coherent condensate \cite{Carusotto2013}. This excitation scheme imposes no stiffness of the condensate phase and thus admits spontaneous formation of the internal polariton flows and the circular currents in the ring geometry. 
We describe the relevant driven-dissipative dynamics in the mean-field treatment using the generalized Gross-Pitaevskii equation coupled to the kinetic equation for the reservoir of incoherent excitons:
\begin{subequations}\label{Eq1}
	\begin{eqnarray}
\notag 	i\hbar \partial _t \Psi&=&\left[-\frac{\hbar ^2}{2m_{\rm pol}}\nabla^2 +g_{c}\left|\Psi\right|^2 + V(\mathbf{r})   \right. \\
		  &+&\left.\left(g_r+\frac{i\hbar s}{2}\right)N_r -  i\hbar\frac{\gamma_c}{2}\right]\Psi, \\
		\partial_t N_r&=&P-\left(\gamma_r + s{\left|\Psi\right|}^2\right)N_r.
	\end{eqnarray}
\end{subequations}
Here $\Psi \equiv \Psi(\mathbf{r})$ stands for the 2D condensate wave function and $N_r \equiv N_r(\mathbf{r})$ for the reservoir density distribution; $m_{\rm pol}$ is the polariton effective mass,  $V(\mathbf{r})$ is the stationary trapping potential. The coefficients $g_c$ and $g_r$ account for the interaction of polaritons between themselves and with the incoherent excitons, respectively. The non-Hermitian terms in (\ref{Eq1}a) describe the dissipation with the rate $\gamma_c$ and the polariton gain $sN_r$ governed by the reservoir density. The shape of the pump $P(\mathbf{r})$ is imprinted on the exciton density distribution whose growth is balanced by the exciton relaxation with the rate $\gamma_r$.

\begin{figure}
	\includegraphics[width=\linewidth]{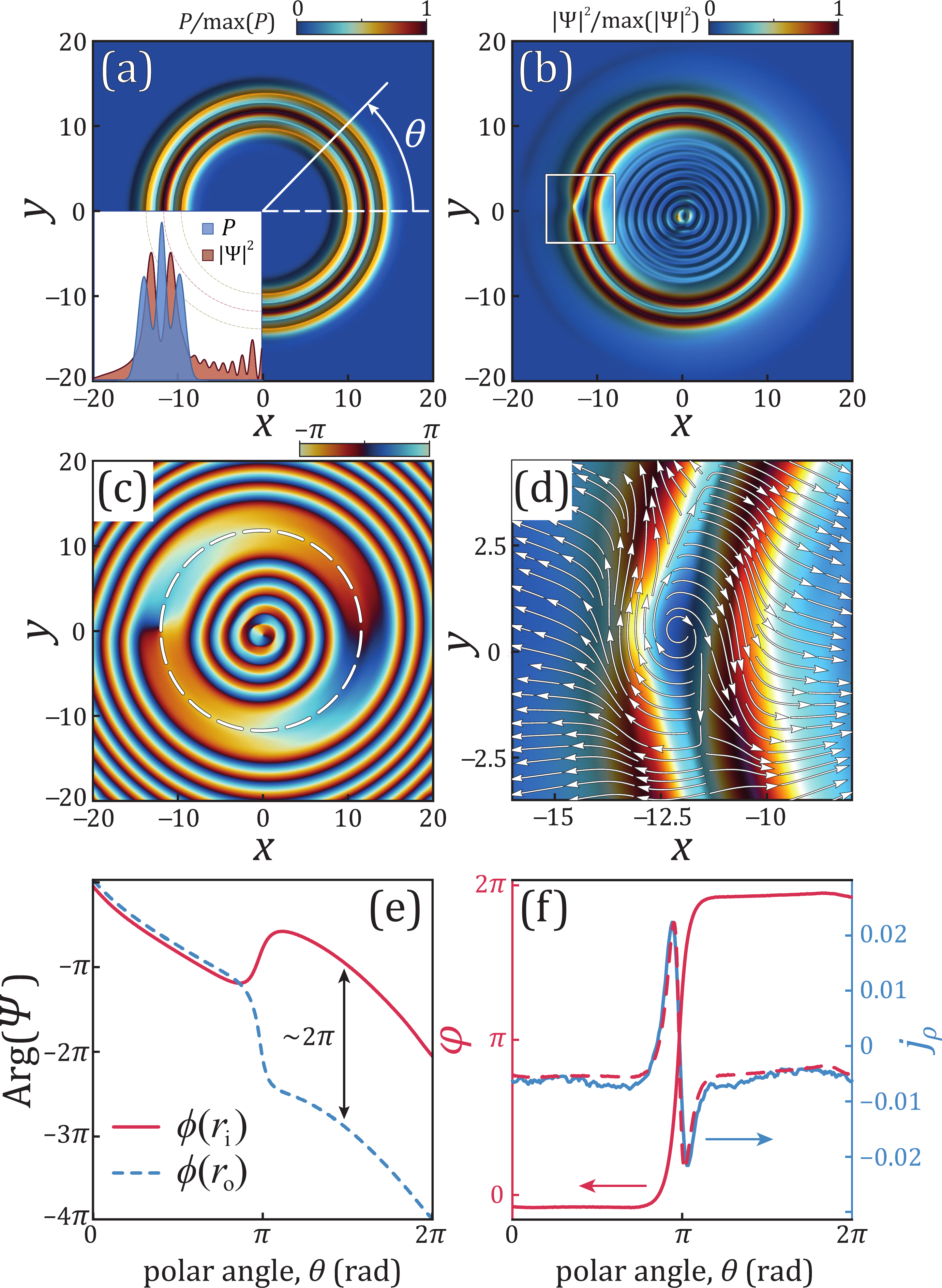}
	\caption{Josephson vortex in the double-ring polariton condensate. (a) The pump distribution. The inset shows the radial cut of the pump (blue) and the polariton density away from the JV (red) distributions. (b) The 2D polariton density pattern. (c) The spatial distribution of the global phase. (d) The vector field of the 2D current density in the region framed in (b). (e) The azimuthal variation of the global phase $\phi={\rm Arg}(\Psi)$ at the potential minima. (f) The relative phase $\varphi$ (on the left, red lines). The radial component of the polariton current in the centre of the junction (on the right, blue lines) shown with the white circle on (c). The solid curve was extracted from the simulations while the dashed one corresponds to the fitting $J=J_c\sin\varphi$ with $J_c=2.25\times10^{-2}$. All the lengths are measured in the units of  $\sqrt{\hbar/(\gamma_c m_{\rm pol})}$, the current in the units  of $\sqrt{\hbar^3\gamma_c^3 /(g_c^2 m_{\rm pol})}$\cite{Note1}.}\label{Fig1}
\end{figure}

Sculpting the pump shape or the trapping potential allows for a precise control of the condensate density distribution. A single ring-shaped condensate can be created using the ring pump \cite{MunozMateo2020}, etching the ring channel \cite{Mukherjee2019} or in the high-$Q$ micropillar cavities \cite{Sedov2021}.  
Creation of the annular Josephson junction thus requires excitation of two concentric condensates. To be specific, we consider the pump beam having a form of three concentric rings, see Fig.~\ref{Fig1}a, which form two annular potential traps for polaritons with the minima's radii $r_{\rm{i}}$ and $r_{\rm{o}}$. We note that the chosen shape of a pump beam can be achieved with use of a spatio-optical modulator \cite{book_digital}. The trapping is mediated by the reservoir excitons which repel polaritons from the pump region. In what follows, the labels ``$\rm{i}$'' and ``$\rm{o}$'' are assigned to the inner and the outer condensates, respectively.

In a single center-symmetric polariton ring, the global circulation of the superfluid is random during the condensation onset. A conventional way to quantify the circulation is using the integer winding number 
\begin{equation}
	m = \frac{1}{2\pi i}\oint_\mathcal{C} \frac{d\Psi}{\Psi}, 
\end{equation}
which is a quantized topological invariant characterizing the phase variation along the closed path $\mathcal{C}$. However in the double-ring geometry, the condensates are mutually dependent. Since the coupling establishes at each point of the junction, the phase locking effect tends to impose the identical azimuthal  profiles of the phase on both rings and thus to equalize their winding numbers. The competition between the coupling of the condensates and the robustness of their individual topological charges will remain the focus of our attention.

The JVs appear as an outcome of this interplay. An example   of the polariton density pattern excited in the double-ring geometry is shown in Fig.~\ref{Fig1}b. The corresponding distribution of the global phase $\phi(\mathbf{r})=\rm{Arg}(\Psi)$ reveals that the inner and the outer condensates are topologically discriminated by the mismatch of their winding numbers, Fig.~\ref{Fig1}c. In the given example, {$m{\rm _i} = -1$ and $m_{\rm o} = -2$}, see Fig.~\ref{Fig1}e. This difference is connected with a single JV located between the rings which is characterized by the $2\pi$ phase winding about its core. 
This is clearly seen in the vector field of the current density shown in Fig.~\ref{Fig1}d and  in the azimuthal dependence of the relative phase $\varphi(\theta)$ which we define as $\varphi=\phi(r_{\rm i}) -\phi(r_{\rm o})$. The phase locking effect imposes synchronous rotation ($\varphi={\rm const}$) everywhere except at the JV position, see Fig.~\ref{Fig1}e. In its vicinity, $\varphi$ experiences a smooth twist on $2\pi$ corresponding to the topological defect akin to the sine-Gordon kink \cite{barone1982,mazo2014}. 

Note that for the condensates tightly confined in the radial direction, $\varphi$ can be associated with the Josephson phase which drives the Josephson current $J = J_c \sin\varphi$.  In the annular junction, the tunnelling current flows in the radial direction. Fig.~\ref{Fig1}d compares the radial component of the polariton current density $j_\rho=(\hbar/m) \Im\left(\Psi^*\partial_\rho\Psi\right)$ at the middle of the junction (the white circle in Fig.~\ref{Fig1}c) with the effective Josephson current evaluated from the numerical data for $\varphi$ using $J_c$ as a fitting parameter 
\footnote{Equations \eqref{Eq1} were solved numerically using the normalized variables. The normalization implies that the time is measured in units of $\gamma_c^{-1}$, the length in units of  $\sqrt{\hbar/(\gamma_c m_{\rm pol})}$, the polariton-polariton interaction strength is rescaled to $1$. The other parameters are $s=1.1 g_c/\hbar$, $\gamma_r=3\gamma_c$, $g_r=2g_c$ and the peak power of the pump is ${\rm max}(P)=5.84 \hbar\gamma_c^2/g_c$.}.  
The asymmetry of  the $j_\rho(\varphi)$-dependence is indicative of the slow motion of the vortex along the annular junction.

\subsection{One-dimensional model of coupled polariton rings}

To get a better insight into the physics standing behind the JV formation, we adopt the effective 1D model of the annular polariton junction developed in \cite{MunozMateo2020}. In what follows we consider the limit of thin rings with large radii which admits reduction of the dimensionality of the problem. In particular,  we use a two-mode ansatz:
\begin{subequations}\label{Eq2}
	\begin{eqnarray}
\label{Eq2a}		\Psi\left(\rho ,\theta \right) &=&{\Psi}_{\rm{i}}\left(\rho ,\theta  \right)+{\Psi}_{\rm{o}}\left(\rho ,\theta \right),
	\\
	\label{Eq2b}	N_r\left(\rho ,\theta\right)&=&N_{\rm{i}}\left(\rho ,\theta \right)+N_{\rm{o}}\left(\rho ,\theta \right),
	\end{eqnarray}
\end{subequations}
where $(\rho ,\theta)$ are the polar coordinates, and assume that the condensates obey central symmetry:
\begin{subequations}\label{Eq3}
	\begin{eqnarray}
		{\Psi}_{\rm{i,o}}\left(\rho,\theta,t\right)&=&Y_{\rm{i,o}}\left(\rho \right)\psi_{\rm{i,o}}(\theta,t), 
		\\
		N_{\rm{i,o}}\left(\rho ,\theta, t \right)&=&Z_{\rm{i,o}}\left(\rho \right)n_{\rm{i,o}}\left(\theta, t \right),
	\end{eqnarray}
\end{subequations}
where the radial functions $Y_{\rm{i,o}}$ are normalized to unity but not orthogonal to each other.  The same is true for the real functions $Z_{\rm i,o}$.

After integrating out the radial dependencies one obtains the following coupled equations \cite{MunozMateo2020}:
\begin{subequations}\label{Eq.1D}
	\begin{eqnarray}
 \notag		i\hbar \partial _t\psi_{\rm{i,o}} &=& \left[-\frac{\hbar^2}{2M_{\rm{i,o}}} \frac{\partial^2}{\partial\theta^2} +\alpha{\left|\psi_{\rm{i,o}}\right|}^2+\alpha_rn_{\rm{i,o}} \right. 
 \\
&+& \left. 		\vphantom{\frac{{\hbar }^2}{2M_{\rm{i,o}}} \frac{\partial^2}{\partial\theta^2}} \frac{i\hbar }{2} \left(r n_{\rm{i,o}} -\gamma\right)\right] \psi_{\rm{i,o}}+\hbar C\psi_{\rm{o,i}}, 
\\
		\partial_t n_{\rm{i,o}}&=&p_{\rm{i,o}}-\left(\Gamma + r\left|\psi_{\rm{i,o}}\right|^2\right)n_{\rm{i,o}},
	\end{eqnarray}
\end{subequations}
which are analogues to the system \eqref{Eq1} except for the linear interaction term governed by the complex coupling parameter~$C$. The explicit expressions for coefficients of the model can be found in \cite{MunozMateo2020}. It is worth mentioning that the kinetic energy associated with the azimuthal motion is small due to the large effective masses $M_{\rm{i,o}} = m_{\rm pol} R_{\rm{i,o}}^2$ which enter the 1D model. For simplicity, we neglect also the difference of the ring radii  assuming $R_{\rm{i,o}}  = R$ and hence $M_{\rm{i,o}}  = M$. Besides, we consider the case of the symmetric pump, $p_{\rm{i}} = p_{\rm{o}} = p$.

The interaction strength $C$ which is a key parameter of the problem is governed by the overlap of wave functions of the condensates located in two rings, which depends on the distance separating them, the pump amplitude and the potential barrier between the rings. 
Taking into account a large flexibility of the considered excitation scheme, we assume that both real and imaginary parts of $C$ can be tuned continuously and independently in a wide range.

Equations \eqref{Eq.1D} have a family of solutions
\begin{subequations}\label{Eq.HomSol}
	\begin{eqnarray}
		\psi_{\rm{i}} &=& \pm \psi_{\rm{o}} = \left( \frac{p}{\gamma  \mp \Im\left[C\right]} - \frac{\Gamma}{r}\right)^{1/2} \exp(im\theta),\\
		n_{\rm{i}}&=&n_{\rm{o}}=\frac{\gamma  \mp \Im\left[C\right]}{r},
	\end{eqnarray}
	\end{subequations}
corresponding to the symmetric and the antisymmetric configurations of the condensates  rotating \textit{synchronously}, i.e. with the same winding number $m$. These states are characterized by locking of the relative phase to $0$ for the symmetric or to $\pi$ for the antisymmetric configuration. Hence they are expected to play a major role in the limit of strong coupling between the condensates.

Those solutions which describe the \textit{asynchronous} rotation of the condensates with different winding numbers must account for the  defects of the relative phase corresponding to the JVs. In the presence of the reservoir and the complex coupling parameter, these solutions can not be found analytically \cite{Kaurov2005,Kaurov2006}. This is why we approach this problem numerically. In the next section, we study the spontaneous nucleation of JVs during the condensate formation.

%%%%%%%%%%%%%%%%%%%%%%%%%%%%%%%%%%%%%%%%%%%%%%%%%%
%%%%%%%%%%%%%%%%%%%%%%%%%%%%%%%%%%%%%%%%%%%%%%%%%%
%%%%%%%%%%%%%%%%%%%%%%%%%%%%%%%%%%%%%%%%%%%%%%%%%%
\section{Spontaneous nucleation of the Josephson vortices}
%%%%%%%%%%%%%%%%%%%%%%%%%%%%%%%%%%%%%%%%%%%%%%%%%%
%%%%%%%%%%%%%%%%%%%%%%%%%%%%%%%%%%%%%%%%%%%%%%%%%%
%%%%%%%%%%%%%%%%%%%%%%%%%%%%%%%%%%%%%%%%%%%%%%%%%%

We solve Eqs.~\eqref{Eq.1D} numerically starting from the weak noise which simulates  fluctuations of the order parameter at the initial stages of condensation. The JVs develop from the phase defects seeded randomly in the initial field distribution. A single JV connects the condensates whose winding numbers differ by one. In the case of $n$ vortices and $n_{\rm a}$ Josephson anti-vortices ($2\pi$ winding of the relative phase is the opposite) trapped between the rings, the winding number quantization condition reads $m_{\rm{i}} - m_{\rm{o}} = n-n_{\rm a}$.

\begin{figure}
	\includegraphics[width=\linewidth]{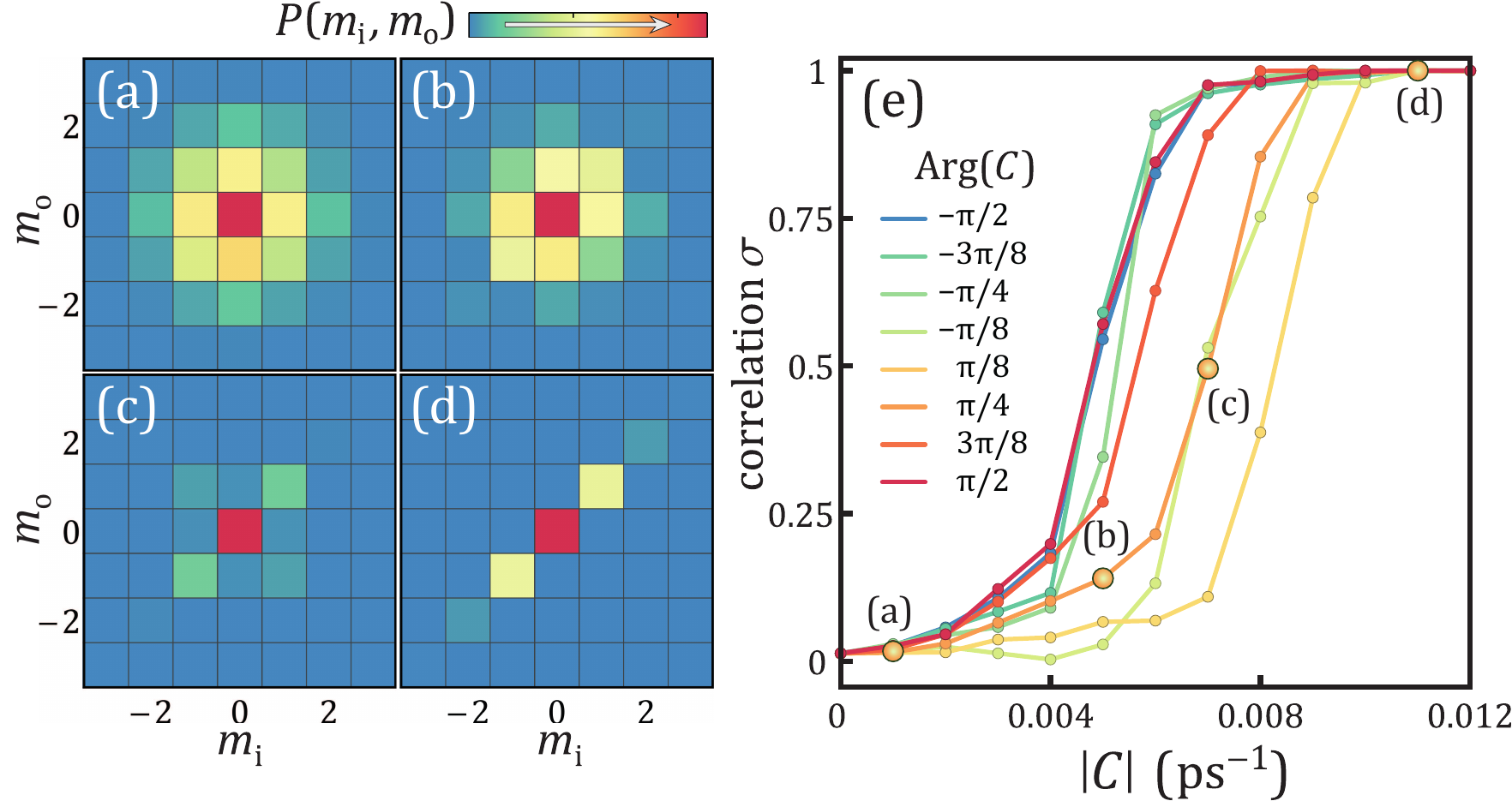}
	\caption{(a)-(d) The histograms of the winding number distribution at the different values of the coupling strength amplitude $|C|$ indicated in the panel (e). The colours of the segments are chosen so that $\sum_{m_{\rm {i}}}\sum_{m_{\rm {o}}}P({m_{\rm {i}}},{m_{\rm {o}}}) = 1$ for each panel. (e) Pearson correlation coefficient  calculated from the data sets of 5000 realizations for each point. The line colour codes the argument $\rm{Arg}(C)$ of the complex coupling parameter. The parameters are $\gamma=0.2$~ps$^{-1}$, $\Gamma=0.15$~ps$^{-1}$, $\alpha/\hbar=0.0025$~ps$^{-1}$, $\alpha_r=2\alpha$, $r=0.02$~ps$^{-1}$, $m_{\rm pol}=10^{-4}m_{\rm fe}$, where $m_{\rm fe}$ is the free-electron mass. The mean radius is $R=15$~$\mu$m, the pump power is $p=2p_{\rm th}$, where $p_{\rm th}=\Gamma\gamma/r$ is the threshold value of the pump in the absence of interaction.}\label{Fig2}
\end{figure}

Each run of the simulations yields a random combination of the winding numbers.  The results of $5000$ runs are summarized on the histograms of the winding number distribution $P(m_{\rm{i}} , m_{\rm{o}})$, Figs.~\ref{Fig2}a-d. Our simulations reveal that the coupling parameter $C$ has a significant impact on the shape of this distribution. To be specific, below we focus on the case of the equal Josephson and dissipative coupling parameters, $\Re[C]=\Im[C]$.

At the weak coupling regime of two rings (small $|C|$), which is typically realized at the large separation distances or high potential barrier,  $P(m_{\rm{i}} , m_{\rm{o}})$ is close to the bivariate normal distribution, see Fig.~\ref{Fig2}a. It implies the absence of any correlations between the inner and the outer rings. In particular, the condensates can rotate either in the same or in the opposite directions with equal probabilities. The weak deviations from the normal distribution appear as the coupling strength increases, Fig.~\ref{Fig2}b. In this case, the synchronous solutions $m_{\rm{i}} = m_{\rm{o}}$ is more favourable than the counterrotating state $m_{\rm{i}} = - m_{\rm{o}}$ whose creation requires excitation of at least $2|m_{\rm{i}}|$ JVs.

With the further increase of the coupling strength the probability of the JV formation falls down. This results in the shrinkage of the distribution along the diagonal occupied by the set of synchronous solutions \eqref{Eq.HomSol} with $m_{\rm{i}} = m_{\rm{o}}$.  The coupling associated with the phase locking effect favours formation of the synchronously rotating condensates and thus tends to destroy the JVs. In this intermediate regime, polariton  currents in the rings are partially correlated. The degree of correlations can be quantified with the use of the Pearson correlation coefficient $\sigma$ 
\begin{equation}
	\sigma(m_{\rm i},m_{\rm o}) = \frac{\sum \left(m_{\rm i} - \bar{m}_{\rm i}\right) \left(m_{\rm o} - \bar{m}_{\rm o}\right)  }   {\sqrt{\sum \left(m_{\rm i} - \bar{m}_{\rm i}\right)^2 \sum\left(m_{\rm o} - \bar{m}_{\rm o}\right)^2 }  },
\end{equation}
where $\bar{m}_{\rm i,o}$ are the mean values and the sums run  over all realizations. The corresponding $\sigma(C)$-dependence is shown in Fig.~\ref{Fig2}e. 

It is important that there is a critical value of $C$ beyond which $\sigma$ approaches $1$ meaning the perfect correlations. The corresponding histogram Fig.~\ref{Fig2}d shows the absence of the asynchronous solutions in this regime, which implies that no JVs survive during the condensate formation.

A gradual increase of the correlation degree shown in Fig.~\ref{Fig2}e 
illustrates the competition between the synchronization effect stemming from the coupling and the topological protection of the circular currents. The phase locking dominates at the strong coupling while in the opposite limit the condensates weakly affect each other. Note that this behaviour remains qualitatively the same at any value of the phase ${\rm Arg}(C)$ of the complex coupling parameter, see Fig.~\ref{Fig2}e. 
In the next section, we focus on the intermediate regime where the interplay between the circular and the Josepshon currents is crucial.

\section{A single Josephson vortex formed between  concentric polariton rings}

At the moderate coupling strength, the presence of JVs has a significant impact on the condensate rotation. We consider the simplest case of a single JV confined in the junction. In particular, we assume that the outer condensate carries no topological charge, $m_{\rm o}=0$, while the inner one spins with $m_{\rm i}=1$. Here we are interested in the stationary solutions neglecting the particular mechanism of the JV formation. 

The interaction between the rotating and the stationary condensates alters the local currents in both rings. The global circulation of the \textit{non-uniform} current  can be characterized by the expectation value of the orbital angular momentum operator $\hat{L}_z=-i\hbar\partial_\theta$. The difference between the mean angular momenta per particle $\ell_{\rm{i,o}}=  N_{\rm i,o}^{-1}\int \psi_{\rm{i,o}}^\ast \hat{L}_z \psi_{\rm{i,o}} d\theta$ of the inner and the outer condensates thus reads
\begin{equation}\label{Eq.Dl}
	\Delta \ell = M\int \left[\frac{j_{\theta\rm i} (\theta)}{N_{\rm i}}  - \frac{j_{\theta\rm o}(\theta)}{N_{\rm o}}    \right] d\theta,
\end{equation}
where $j_{\theta\rm i,o} = (\hbar/M) \Im\left( \psi_{\rm i,o}^*\partial_\theta\psi_{\rm i,o} \right)$ quantifies the azimuthal component of the current density in the inner (outer) ring, $ N_{\rm i,o} = \int |\psi_{\rm{i,o}}|^2 d\theta$ are the populations of the condensates. Note that for the uniform states $\psi_{\rm i,o}= \sqrt{\rho_{\rm i,o} } \exp\left(i m_{\rm i,o}\theta \right)$ which are realized at $C=0$, the average angular momenta coincide with the winding numbers up to the dimensional constant, $\ell_{\rm i,o}= \hbar m_{\rm i,o}$.
%the immobile ring.

\begin{figure}
	\includegraphics[width=\linewidth]{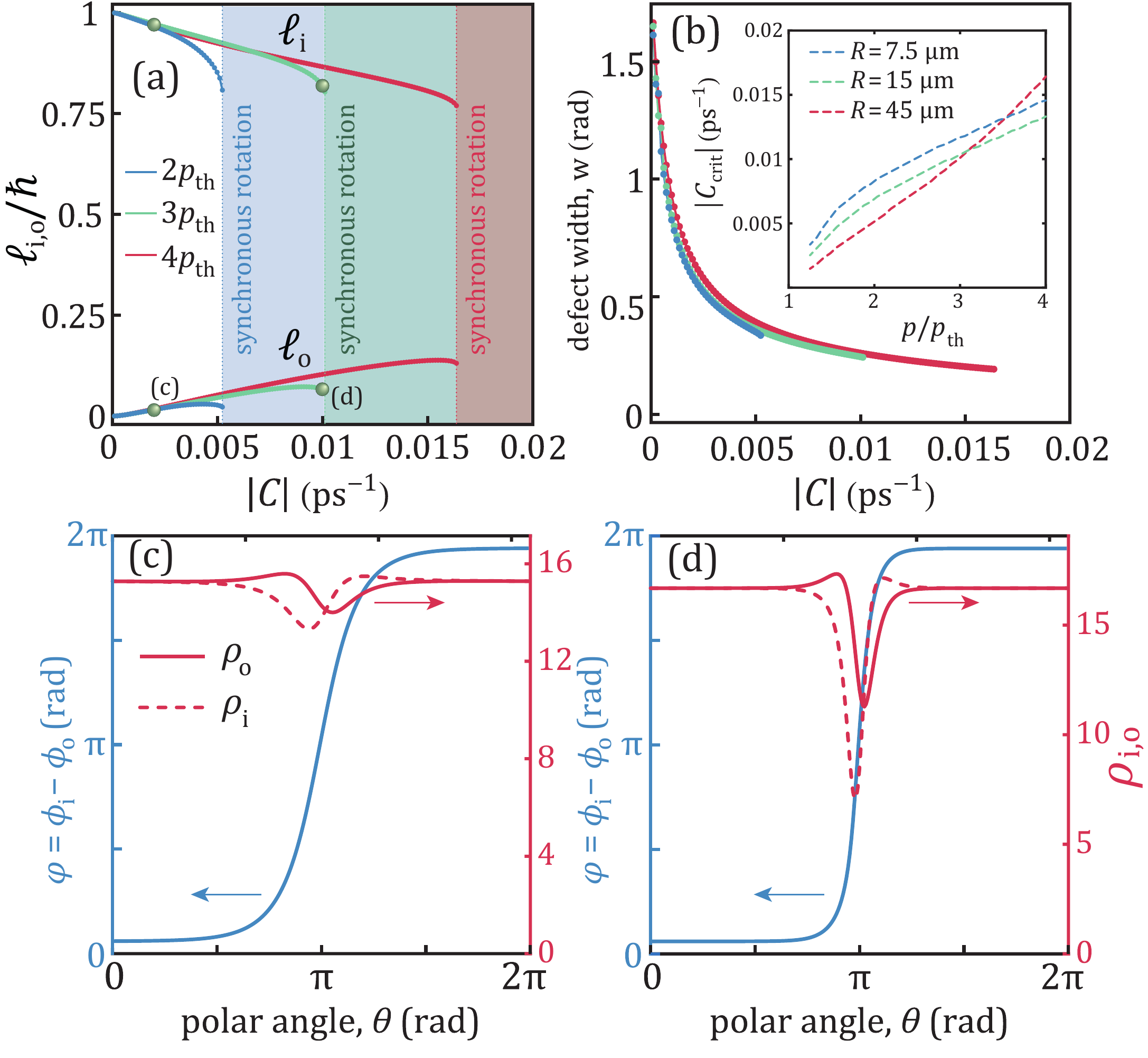}
	\caption{(a) The dependence of angular momenta of the inner and outer rings on the coupling strength at three different values of the pump power indicated on the panel.  (b) The effective width $w=\sqrt{\int\theta^2 \partial_\theta\varphi d\theta - \left(\int\theta \partial_\theta\varphi d\theta\right)^2}$ of the JV at the same values of the pump power as in (a). The meaning of the colours remains the same as in (a). The inset shows the critical value of the coupling strength at three different values of the mean radii of the rings.  (c) and (d) show  the relative phase $\varphi(\theta)$ together with the densities profiles $\rho_{\rm i,o}(\theta)$ at (c) $C=0.002$~ps$^{-1}$ and (d)  $C=0.01$~ps$^{-1}$. For all panels $R=45$~$\mu$m and $\Re(C) = \Im(C)$.}\label{Fig3}
\end{figure}

At the finite coupling strength, the interaction of two ring condensates speeds up the condensate in the immobile ring so that its angular momentum increases. At the same time, the rotation of the spinning ring slows down as if it would be affected by the drag force. The effect of this pseudo-drag is shown in Fig.~\ref{Fig3}a. The angular momenta vary continuously approaching each other as the coupling strength increases. The resulting difference of the angular momenta $\Delta \ell$  is governed mainly by the current circulation about the JV core according to the definition \eqref{Eq.Dl}. Indeed, assuming $N_{\rm i} \approx  N_{\rm o} = N$ under symmetric excitation, one can write $\Delta \ell = (\hbar/N) \int \left[ \rho_{\rm i}(\theta) \partial_\theta \phi_{\rm i} - \rho_{\rm o}(\theta) \partial_\theta \phi_{\rm o} \right] d\theta$. Away from the JV, the rotation of the condensates is synchronized locally, $\rho_{\rm i}(\theta) = \rho_{\rm o}(\theta)$ and $\phi_{\rm i}(\theta)$  = $\phi_{\rm o}(\theta)$, -- see Fig.~\ref{Fig3}c,d. Hence this region makes no contribution to $\Delta \ell$. 

\begin{figure}%[h!]
	\includegraphics[width=\linewidth]{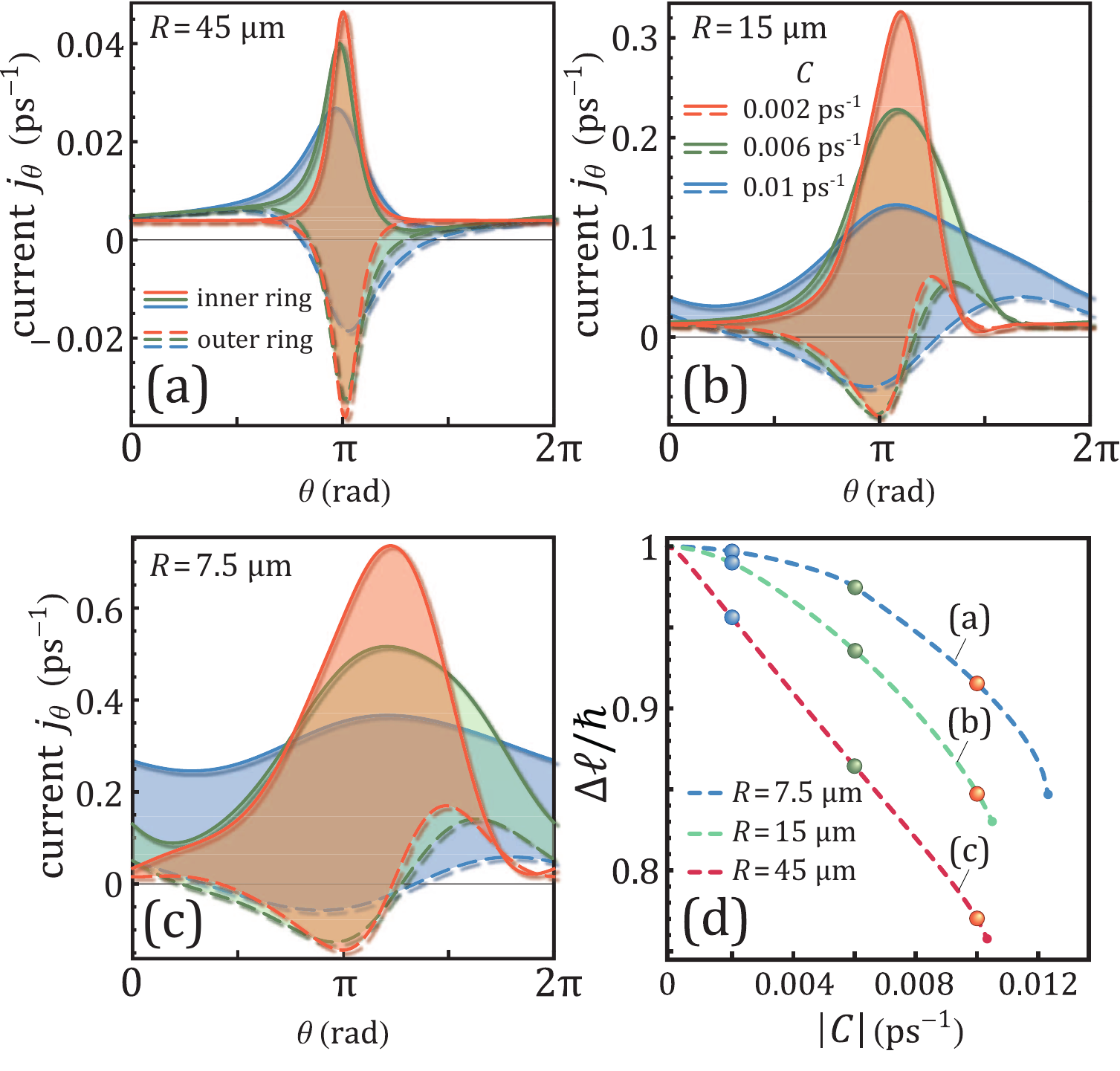}
	\caption{The variation of the azimuthal density currents for the inner (solid lines) and the outer rings (dashed lines) for (a) $R=45$~$\mu$m, (b) $R=15$~$\mu$m and (c) $R=7.5$~$\mu$m. The blue lines correspond to $C=0.002$~ps$^{-1}$, the green ones to $C=0.006$~ps$^{-1}$ and the orange ones to $C=0.01$~ps$^{-1}$. The areas of the shaded regions govern the difference of the average angular momenta $\Delta\ell$. (d) The dependence of $\Delta\ell$ on the coupling amplitude for three different values of the mean ring radius $R$ corresponding to the panels (a), (b) and (c). The particular cases shown in panels (a)-(c) are indicated with dots of matching colours. For all panels $p=3p_{\rm th}$ and $\Re(C) = \Im(C)$.
	}\label{Fig4}
\end{figure}

The azimuthal distribution of the density currents is shown in Fig.~\ref{Fig4}a. The main contribution to the angular momenta difference comes from the region  where the currents flow in opposite directions. This region stems from the $2\pi$-winding of the relative phase $\varphi$ which is concomitant of the JV. As the coupling strength grows, the azimuthal size of the JV decreases, see Figs.~\ref{Fig3}c,d. Since the phase-winding region becomes more steep at stronger coupling, the amplitudes of the opposite currents increase, see Fig.~\ref{Fig4}a. The competition of these processes results in the decrease of the angular momentum difference, see Fig.~\ref{Fig4}d. 

One can conclude that the efficiency of the pseudo-drag effect between the rings correlates with the vortex structure, especially with its size. The physical scale of the JV can be estimated as  $\lambda =  \hbar   / \sqrt{2 M |C|} $ by analogy with the Josephson penetration depth which is a characteristic length scale of the superconducting Josephson junction \cite{barone1982}. The inverse $\lambda(C)$-dependence fits well the results of our simulations. Fig.~\ref{Fig3}b shows the calculated value of the width $w$ of the JV quantified by the variation of the relative phase $\partial_\theta\varphi$ which reflects the steepness of the phase defect. 

Note that the defect width $w$ demonstrates a weak dependence on the pump power. However, the impact of the ring radii on it is significant. Since  $M\propto R^2$ the Josephson penetration length $\lambda$ diverges at small radii, $\lambda \propto R^{-1}$. This effect is seen in Figs.~\ref{Fig4}b,c. Even at the moderate value of the radius $R=15$~$\mu$m, the width of the JV is comparable with the circumference of the junction.  With the further decrease of the radii, the JV starts occupying the entire annular junction. The growth of the JV leads to the decrease of the angular momentum difference which implies suppression of the pseudo-drag effect, see Fig.~\ref{Fig4}d.

%Note: The mismatch in the effective masses $M_i \neq M_o$ prohibits synchronization, i.e. the synchronous states appear at larger values of the coupling strength (the synchronization threshold increases or shifts towards larger values of the coupling strength). Besides, it increases the velocity of the defect rotation.

\section{Synchronization via phase slip}

The dependencies shown in Figs.~\ref{Fig3}a and \ref{Fig4}d demonstrate the presence of the critical value of the coupling strength $C_{\rm crit}$ which corresponds to the collapse of the Josephson vortex. This behaviour manifests the domination of the phase locking effect in the strong coupling regime where rotation of the condensates is fully synchronized. Since the polariton JV disappears abruptly  in contrast to the JV in the atomic condensates \cite{Kaurov2005,Kaurov2006}, this phenomenon should be associated with the loss of stability of this state. The synchronization occurs via the phase slip event which corresponds to the vortex crossing either of the rings. In the considered case, if the vortex enters the inner region, it compensates the topological charge of the inner ring decreasing it by one, $m_{\rm i} = 0$. In the opposite case the JV  crosses the outer ring and increases its winding number living the condensates in the state $(m_{\rm i},m_{\rm o}) = (1,1)$.

So far, we considered the case of the equal Josephson and dissipative coupling parameters. Now we shall discuss the role of the phase of the complex coupling ${\rm Arg} (C)$. In particular, we focus on the behaviour of the critical coupling strength $C_{\rm crit}$. 

\begin{figure*}
	\includegraphics[width=0.72\linewidth]{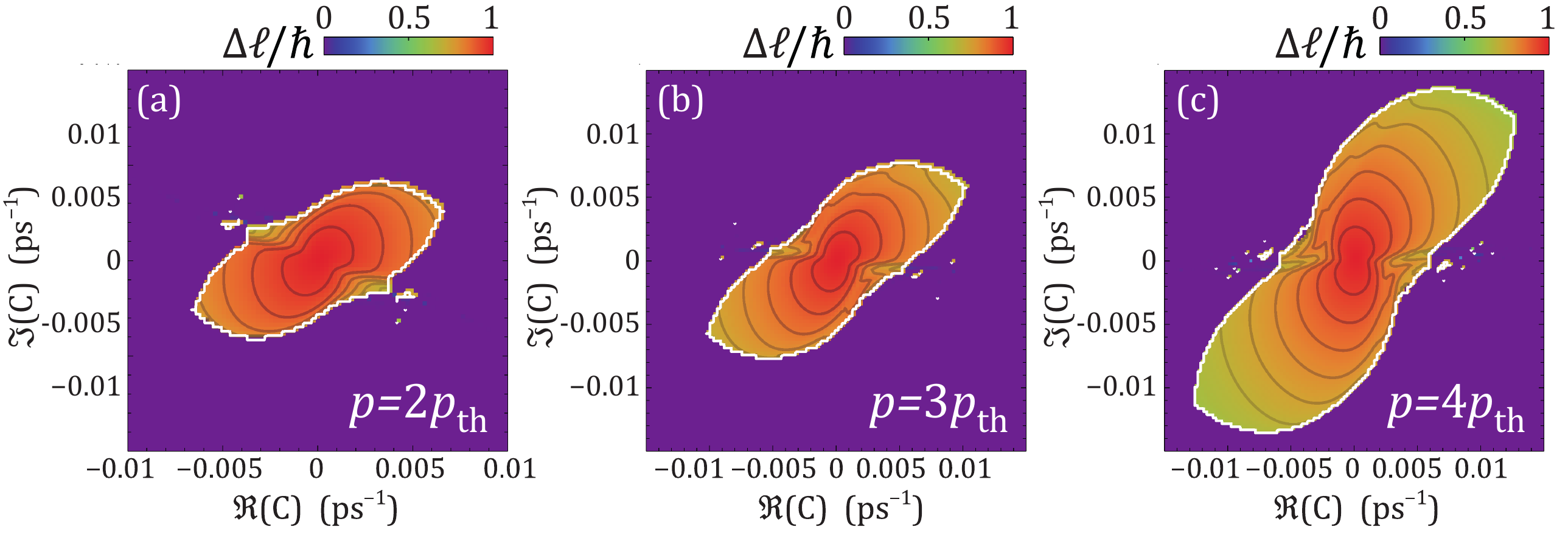}
	\caption{Synchronization maps on the parameter space $\left(\Re(C),\Im(C)\right)$ for various values of the pump power: (a) $p=2p_{\rm th}$, (b) $p=3p_{\rm th}$ and (c) $p=4p_{\rm th}$, where $p_{\rm th}=\Gamma\gamma/r$. The mean ring radius is $R=15$~$\mu$m.}\label{fig5}
\end{figure*}

Fig.~\ref{Fig3}a shows that the value of $C_{\rm crit}$ depends crucially on the pump power. The strong pump stabilizes the JV shifting position of the critical point towards stronger couplings. The similar scenario occurs at any ${\rm Arg} (C)$ as it is demonstrated in Fig.~\ref{fig5} which shows the synchronization maps on the complex plane $\left(\Re(C),\Im(C)\right)$ at different pump intensities. As in the previous section, it is assumed that $m_{\rm o}=0$ and $m_{\rm i}=1$ in the asynchronous state. These maps show the gradual decrease of the mean angular momenta difference from $\Delta\ell=1$ at $C=0$ to the minimal value at the phase boundary indicated by the white line. Beyond this line, the synchronous solution with $\Delta\ell=0$ establishes. 

The phase boundary demonstrates a significant dependence on the ${\rm Arg} (C)$. Note that each distribution is symmetric with respect to the rotation by $\pi$. It means that the synchronization scenario depends on whether the real and imaginary parts of the coupling parameter are of the equal or the opposite signs.

\section{Conclusions}

Polariton condensates excited in the double-ring geometry are able to rotate either in the same direction or in opposite directions and carry arbitrary topological charges. The mismatch of the winding numbers of two condensates lead to the formation of Josephson vortices within the annular junction. The coupling between two rings affects their rotation.  Due to the interplay between the local circular currents inherent in the driven-dissipative condensates and the radial tunnelling current which favours locking of the relative phase, the rotating rings demonstrate the behaviour akin to the drag effect. We predict that at coupling strengths exceeding a critical value dependent on the radii of the rings the Josephson vortex would decay through a phase slip event.

\begin{acknowledgments}
	This work is supported by the Westlake University (Project No.~041020100118) and from Program 2018R01002 funded by the Leading Innovative and Entrepreneur Team Introduction Program of Zhejiang. I.C. acknowledges funding from National Natural Science Foundation of China (Grant No. 12050410250). The support from RFBR grant 21-52-10005, from the Grant of the President of the Russian Federation for state support of young Russian scientists No. MK-5318.2021.1.2 and from the state task  in the scientific activity project 0635-2020-0013 is acknowledged. I.A.S. acknowledges support from Russian Foundation for Basic Research (RFBR), in framework of the joint RFBR-DFG project No. 21-52-12038. A.K. acknowledges the support from the Road Map for Quantum Computing program of the Rosatom.
\end{acknowledgments}

%\bibliography{APFBibl}
%

\end{document}